\documentclass[aps,prl,showpacs,showkeys,preprint,groupedaddress]{revtex4}

\usepackage[dvips]{graphicx}
\usepackage{amssymb}
\usepackage{amsthm}
\theoremstyle{definition}
\newtheorem{dfn}{Definition}[]
\newcommand{\bm}[1]{\mbox{\boldmath $#1$}}

\begin{document}


\title{Leptokurtic Portfolio Theory}


\author{Robert Kitt}
\email[]{kitt@ioc.ee}
\affiliation{Institute of Cybernetics at Tallinn University of Technology}
\affiliation{Hansa Investment Funds, Liivalaia 12, 15038, Tallinn, ESTONIA}

\author{Jaan Kalda}

\affiliation{Institute of Cybernetics at Tallinn University of Technology, Akadeemia tee 21, 12618, Tallinn, ESTONIA}


\date{\today}

\begin{abstract}
The question of optimal portfolio is addressed. The conventional Markowitz portfolio optimisation is discussed and the shortcomings due to non-Gaussian security returns are outlined. 
A method is proposed to minimise the likelihood of extreme non-Gaussian drawdowns of the portfolio value. 
The theory is called Leptokurtic, because it minimises the effects from "fat tails" of returns. The leptokurtic portfolio theory provides an optimal portfolio for investors, who define their risk-aversion as unwillingness to experience sharp drawdowns in asset prices. Two types of risks in asset returns are defined: a fluctuation risk, that has Gaussian distribution, and a drawdown risk, that deals with distribution tails. These risks are quantitatively measured by defining the "noise kernel" -- an ellipsoidal cloud of points in the space of asset returns.
The size of the ellipse is controlled with the threshold parameter: the larger the threshold parameter, the larger return are accepted for investors as normal fluctuations. 
The return vectors falling into the kernel are used for calculation of fluctuation risk. Analogously, the data points falling outside the kernel are used for the calculation of drawdown risks. As a result the portfolio optimisation problem becomes three-dimensional: in addition to the return, there are two types of risks involved. Optimal portfolio for drawdown-averse investors is the portfolio minimising variance outside the noise kernel. The theory has been tested with MSCI North America, Europe and Pacific total return stock indices. 
\end{abstract}


\maketitle

\section{Introduction}
The bridges between statistical physics and financial economics have been recently crossed by many of authors, cf.~\cite{mantegna00,voit01,roehner02,bouchaud03}. The topics of research include the descriptive statistics of the price movements, market microstructure models and many others. Recently the problem of basket of assets has attracted attention (cf.~\cite{bouchaud03,muzy01}). The portfolio optimisation has been introduced in 1950s by Harry Markowitz \cite{markowitz91}. The simplicity of portfolio optimisation problem has made it well-accepted in financial community (cf.~\cite{elton95} and references therein). The conventional or Markowitz portfolio theory (MPT) assumes the Gaussian probability distribution function for security returns -- this is widely questioned in Econophysics literature. The recent reports show that even the L\'evy stable distributions are not describing the stochastic process of price changes \cite{gabaix03}. 
Furthermore, the temporal organisation of the price increments is also complicated, multifractal, c.f.\ \cite{mantegna00}.
Apparently, the optimal portfolio question should be reconsidered bearing in mind these very important findings.
However, these revised methods have to be simple and robust. Indeed, the statistical data available for the analysis are typically 
insufficient for a statistically meaningful application of advanced and complicated techniques, 

In this paper the theory is provided that leads to three-dimensional portfolio optimisation. In order to keep the approach as simple as possible, we ignore the multifractal aspects of the price movements.
The risk of the portfolio is split to two: Gaussian risks and ``fat tail'' risks. Then, the portfolio choice is provided for the investors who want to minimise the absolute drawdowns. In this paper the Leptokurtic Portfolio Theory is introduced and motivated. Further, the application is provided. Finally, the approach is tested based on various international stock indices. 

\subsection{Definitions}
Let $r_i(t,\Delta t)$ be the return of the security $i$ in the portfolio at time $t$ for recent period of $\Delta t$:
\begin{equation}
\label{return}
r_i(t,\Delta t)\equiv \ln p_i(t)-\ln p_i(t-\Delta t)\simeq\frac{p_i(t)-p_i(t-\Delta t)}{p_i(t-\Delta t)}
\end{equation}
where $p_i(t)$ denotes the price of security $i$ at time $t$. As noted earlier, the statistical properties of quantity $r(t)$ are well-elaborated in the literature. MPT assumes the Gaussian distribution -- an obvious disagreement with recent findings \cite{gabaix03}. The portfolio return -- simple sum of the returns of portfolio constituents and portfolio risk --  standard deviation $\sigma_p$ -- are defined as follows:
\begin{equation}
\label{eq:portf-return}
r_p(t,\Delta t)=\sum_{i=1}^nw_ir_i(t,\Delta t)
\end{equation}

\begin{equation}
\label{risk}
\sigma_p^2=\sum_{i=1}^n \sum_{j=1}^n w_i w_j\sigma_{i,j}
\end{equation}
where $\sigma_{i,j}$ is the covariation of returns of securities $i$ and $j$; $\sigma_{i,i}\equiv\sigma_i^2$. Under the assumption of Gaussian returns, the measured historic values of returns and covariations can be used as a proxy to the future. 

The expected (i.e. {\it ex ante}) return in time-window $\tau$ is defined as average of previous realisations:
\begin{equation}
\label{eq:return}
\eta_i=E(r_i)=\left<r_i\right>_{\tau}
\end{equation}
The covariance of returns for security pair is found as follows: 
\begin{equation}
\label{eq:covar}
\sigma_{i,j}=\left<r_ir_j\right>_{\tau}-\left<r_i\right>_{\tau}\left<r_j\right>_{\tau}=\left<r_ir_j\right>_{\tau}-\eta_i\eta_j
\end{equation}
If $i=j$ then the covariance relaxes to simple variance: $\sigma_{i,i}\equiv \sigma_i^2=\left<r_i^2\right>_{\tau}-\eta_i^2$.

It is convenient to represent the portfolio in the matrix form. Denoting the return and weight vectors respectively with $\eta$ and $w$, Eq.~(\ref{eq:portf-return}) is rewritten as
\begin{equation}
r_p=\eta^T w,
\end{equation}
where $\eta^T$ is transpose of $\eta$. Further, let {\bf C} be the matrix of covariation coefficients with elements of $\sigma_{i,j}$. The Eq.~(\ref{risk}) becomes:
\begin{equation}
\label{eq:risk-matrix}
\sigma_p^2=w^T{\bf C}w
\end{equation}

\section{Leptokurtic portfolio theory}
What is the quantitative measure of a portfolio risk in the case of strongly non-Gaussian price fluctuations with (possibly) infinite variance and L\'evy-like spectrum of price jumps? Note that stable L\'evy distribution of price jumps assumes diverging variance; the power law reported by X.~Gabaix et al \cite{gabaix03} assumes diverging kurtosis (fourth moment).
In the case of diverging fourth moment, the distributions are called {\em leptokurtic}. The tails of the leptokurtic distributions are "fatter" than predicted by Gaussian distribution. In fact, all power-law distributions lead to leptokurtic distributions and that must be accounted by determination of risk. The obvious question arises: what is the risk of portfolios that obey leptokurtic distribution tails? Here, a model is proposed to offer the portfolio choice under such assumptions. First, the two types of fluctuations are separated: the fluctuation or Gaussian risk (which can be called a "good" risk, because the distribution tails approach quickly zero) and the drawdown, or power-law risk ("bad" risk, because the actual portfolio loss can be much larger than that of predicted by Gaussian approach). These two types of risk, together with the expected return, create a three-dimensional space; second subsection is devoted to the discussion of the portfolio optimisation in that space. Finally, a simple empirical illustration is provided using international stock indices.

\subsection{Separation of risks}
The power-law distribution of security returns leads to a large amount of "small" price movements, and few "large" movements. 
It is convenient to separate the "noise kernel" of daily returns --- these are the returns which are smaller than few standard deviations.
First, let us assume that the volatilities are not diverging.
Then, the "noise kernel" is defined as a set of points $Q$ in the $n$-dimensional space of returns
(where $i$-th axes measures the return $r_i$ of the $i$-th asset) falling
into such an ellipsoid where the probability density function of return vectors is above a (small) threshold.
This threshold serves as a model parameter (see below).
Using the eigenvectors of the covariance matrix as the orthonormal basis $\rho_i$, the ellipsoid is defined as
\begin{equation}
\label{eq:noise-kernel0}
Q\in \sum_{i=1}^n\rho_i^2\lambda_i^{-1} \le n\theta, 
\end{equation}
where $\lambda_i$ is the $i$-th eigenvalue of the covariance matrix.
The threshold parameter $\theta \sim 1$ should be chosen in such a way that the probability for a return vector 
falling outside the ellipsoid, is of the order of few percents; it 
regulates the ellipsoid "size" in units of standard deviations. In the case of two assets, the ellipsoid becomes ellipse, and can be expressed as
\begin{equation}
\label{eq:noise-kernel}
Q\in \tilde r_i^2\sigma_j^2+\tilde r_j^2\sigma_i^2-2\tilde r_i\tilde r_j\sigma_{i,j} \le \left[2(\sigma_i^2\sigma_j^2-\sigma_{i,j}^2)\right]\theta
\end{equation}
Hereinafter, $\tilde r$ denotes the zero-shifted return, i.e. $\tilde r_i = r_i-\eta_i$. The shift is necessary to keep the centre of the ellipse in the 0-point of the coordinates. 

For diverging volatilities, the above outlined approach can be still applied; however, the net volatility 
is then a non-stationary quantity and depends on the particular observed set of  extreme price movements. The effect of these extreme movements can be removed by an iterative approach: instead of using the net 
covariance matrix for the calculation of the noise kernel, the noise kernel's own  covariance matrix  must be used.

The investors are looking for capital appreciation, in order to achieve the investment performance targets. What the investors do not like, is fluctuation in asset prices. Indeed, everybody would be happy to see a linear appreciation of capital with a suitable growth rate. However, investors would accept the "normal" fluctuations with a low standard deviations. What the investors really want to avoid, is a sharp drawdown in asset prices. Thus, the risk components should be separated to Gaussian ("good") fluctuation risk, and drawdown risk. 
This is done as follows: 

\begin{dfn}
\label{def:fluct-risk}
{\bf Fluctuation risk}: The risk is measured as the average deviation from average return, i.e. risk is measured with standard deviation. 
\end{dfn}

\begin{dfn}
\label{def:draw-risk}
{\bf Drawdown risk}: The risk is measured as the likelihood of sharp changes in asset prices. Risk is the degree of willingness to experience sharp moves in asset prices. 
\end{dfn}

According to MPT, the drawdown risk in not existing. This is actually not true, if one thinks back to the history: stock market crashes have occurred quite frequently: 1929 and 1987, 2000 in US, 1997 South-East Asia, 1998 Russia are just some well-known examples. In order to separate the risks, the measure of correlation should be revised. Here, the concept of {\em exceedance correlation} (cf.~\cite{longin98,longin01,cizeau01}) is extended. The idea of separation of risks was also suggested by Chow {\it et al} (cf.~\cite{chow99}).

Our simplified approach is as follows. For the calculation of the fluctuation risk of the portfolio (Def.~\ref{def:fluct-risk}), 
only these data points which are in the noise kernel are accounted for. 
On the other hand, the points outside noise kernel are used for the calculation of drawdown risk of the portfolio, and of the related covariance matrix of returns (Def.~\ref{def:draw-risk}). 
Note that the 
larger the threshold parameter $\theta$, the larger fluctuations are considered as "normal" or "acceptable". 
To conclude, equations  

\begin{equation}
\label{eq:correl-fluct}
\sigma_{i,j|\mbox{fluc}}(\theta)=\left<\tilde r_{i}\; \tilde r_{j}\right>_{\bm{r}\in Q}-\left<\tilde r_{i}\right>_{\bm{r}\in Q}\left<\tilde r_{j}\right>_{\bm{r}\in Q}
\end{equation}
and
\begin{equation}
\label{eq:correl-draw}
\sigma_{i,j|\mbox{draw}}(\theta)=\left<\tilde r_{i}\; \tilde r_{j}\right>_{\bm{r}\ni Q}-\left<\tilde r_{i}\right>_{\bm{r}\ni Q}\left<\tilde r_{j}\right>_{\bm{r}\ni Q}
\end{equation}
are used to calculate the covariation coefficients for fluctuation risk and drawdown risk respectively. Here, $\left<\cdots\right>_{\bm{r}\in Q}$ (and
$\left<\cdots\right>_{\bm{r}\ni Q}$) denote averaging over all the return vectors $\bm r$ belonging (and not belonging) to the kernel $Q$.
The covariation matrix {\bf C} used in MPT [cf.~Eq~(\ref{eq:risk-matrix})] is thus split into ${\bf C}_{\in Q}$ and ${\bf C}_{\ni Q}$, corresponding to the fluctuation and drawdown regimes, respectively.

\begin{figure}[loc=htbp]
\caption{Exclusion of the noise kernel from covariation analysis. a) Two normalised, but correlated (with $\rho\simeq 70\%$) Gaussian variables; b) daily returns of Standard \& Poors's 500 and DAX equity indices.}
\includegraphics[scale=.6]{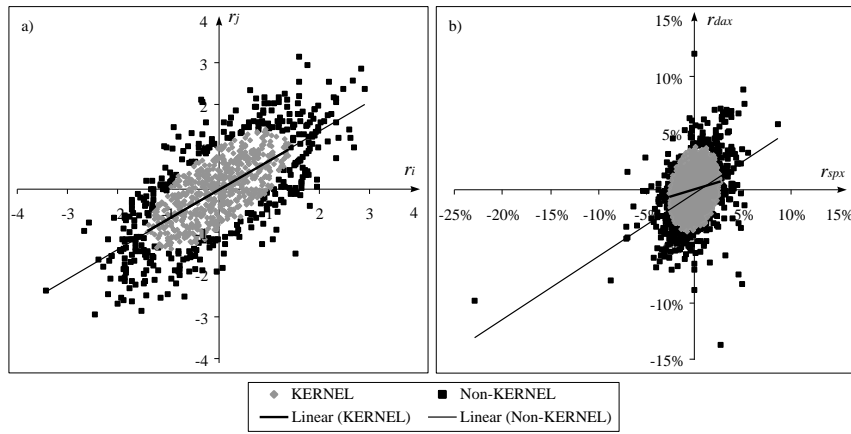}
\label{fig-nonoise}
\end{figure}

The separation of noise kernel is shown schematically in Fig.~\ref{fig-nonoise}, where the data points and their linear regression lines are plotted. 
In the left panel of Fig.~\ref{fig-nonoise}, two normalised (with zero mean and unit standard deviation), but correlated ($\rho\simeq 70\%$) random Gaussian variables are plotted. The noise-kernel corresponds to the threshold parameter $\theta=1$. We are dealing with a pure Gaussian distribution; therefore, the correlation does not change with the exclusion of the noise kernel. 
A different behaviour is observed in the right panel of Fig.~\ref{fig-nonoise}. There are daily returns plotted for US Standard \& Poors' 500 (spx) and German Deutsche Aktieindex (dax) equity indices. The period ranged from September 1959 to March 2004 (i.e. total more than 11000 data points) and the kernel was defined with parameter $\theta=3$. As seen, the noise kernel and the rest of the data have quite different correlations.

\subsection{Application of LPT}
Leptokurtic Portfolio Theory (LPT) is a simple generalisation of MPT. 
Equations (\ref{eq:noise-kernel}) and (\ref{eq:correl-draw}) yield non-trivial results only when large asset price jumps (drawdowns) exist. The boundary between the ordinary fluctuation risk and extraordinary drawdown risk is defined by the parameter $\theta$. Here, we have presented market 
data analysis with $\theta=3$; the value $\theta=2$ yields similar results.
Note that in the case of Gaussian distribution, the events fall into three standard deviations with probability equal to $99.73\%$, which seems to be a reasonable  crossover point. 

The application of LPT introduces additional dimension to the Gaussian risk return space: non-Gaussian risks. So, the portfolios are evaluated in three-dimensional space containing two types of risks, and the return. 
This allows us to separate the fluctuation and drawdown risks and the portfolio choice gets more complicated. 
Investors must choose an optimal balance between the Gaussian risk and return, but they have also to evaluate the potential of the "out-of-statistics" drawdowns. To elaborate this concept, consider the example 
of Fig.~\ref{fig-nonoise} b). 
Suppose the investor is willing to invest into a combination of the indices of SPX and DAX. If the investor is accepting the fluctuations within three standard deviations, but he is unwilling to accept the large drawdowns, then he should invest into a portfolio with minimised drawdown risk. On the other hand, if the low values of the average short-term Gaussian fluctuation is 
more highly prioritised than the desire to minimise the likelihood of drawdowns, minimised kernel-risk portfolio satisfies the investment goal. 

Equation~(\ref{eq:return}) provides the definition of returns in MPT. In the spirit of LPT, one could wish to distinguish between "ordinary" (Gaussian) and "extraordinary" returns. However, in the case of predicted returns, such splitting seems unjustified: first of all, because the use of historic returns as a proxy to the future returns is far from being justified. Indeed,
due to the non-stable nature and non-stationarity of returns, there will be large differences between short-term realisations. This will inevitably lead to the fact that historic returns do not provide the valid forecast to the future returns -- a circumstance that is easy to check in any financial time series. In fact, the expected return of the security depends also on the choice of investment horizon -- it varies from investor to investor. To conclude, we leave the expected return un-splitted and prefer more fundamental (i.e. economic result based) approach in estimating the future returns.

\subsection{Empirical evidence: the validity of LPT}
\begin{figure}[loc=htbp]
\caption{Determination of minimum risk portfolios of MSCI North America and MSCI Pacific time series.}
\includegraphics[scale=.6]{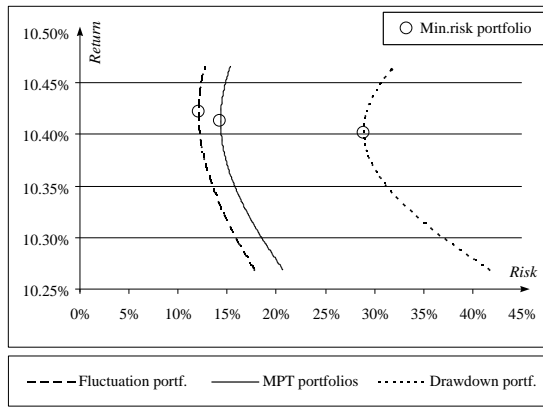}
\label{fig-LPT}
\end{figure}

In previous, we have defined the idea of separation of risks and their measures. Here, the empirical evidence is provided. We construct the portfolios by using monthly data of the following Morgan Stanley Capital International (MSCI) total return indices: MSCI Europe (hereinafter denoted as {\em E}), MSCI North America ({\em NA}) and MSCI Pacific ({\em P}) 
in the period from December 1969 to February 2005 (total more than 420 months). In this paper, we only look at the examples of pairs of securities. The portfolio problem with $N$ securities will be addressed in future. The method used is the following: 
\begin{enumerate}
	\item we take the pair of securities and find the noise kernels with $\theta=1,2,3$ according to Eq.~(\ref{eq:noise-kernel}),
	\item we determine the kernel and non-kernel covariations according to Eqs.~(\ref{eq:correl-fluct}) and (\ref{eq:correl-draw}),
	\item we determine the minimum-risk portfolios according to fluctuation, drawdown and Markowitz definition (the latter is simply aggregated standard deviation of kernel and non-kernel) with parameter $\theta=1,2,3$,
	\item we back-test the minimum-risk portfolios with the same data-set by investing hypothetical unit currency in December 1969
\end{enumerate}
This process is illustrated with an example of MSCI North America and MSCI Pacific time series. In Table~\ref{table:NAP}, the correlation matrices are presented for MPT, Fluctuation, and Drawdown methods using the value $\theta=3$. In Fig.~\ref{fig-LPT}, the portfolio choice is shown using the same data as in Table~\ref{table:NAP}. The risk-return space in Fig.~\ref{fig-LPT} is constructed in a spirit of MPT: in the abscissa, the risk in annualised standard deviations is plotted. Note that for different portfolio sets, the different definitions of covariations and standard deviations are used. The portfolios with minimum variance are marked with circles.

\begin{table}[loc=htbp]
\caption{Covariation matrices of MSCI North America and MSCI Pacific covariations}
\label{table:NAP}
\begin{tabular}{|p{1.2cm}|*{3}{p{1.5cm} p{1.5cm}|}}
\hline
&MPT&&Fluctuation&&Drawdown&\\ \hline\hline
$\sigma_{NA,P}$&NA&P&NA&P&NA&P\\ \hline
NA&0.1959\%&0.1037\%&0.1365\%&0.0797\%&0.8648\%&0.3427\%\\ \hline
P&0.1037\%&0.3563\%&0.0797\%&0.2647\%&0.3427\%&1.4603\%\\ \hline
\end{tabular}
\end{table}

Finally, a simple back-test is performed with portfolios found as follows: a hypothetical unit of currency is invested into optimal portfolios found in previous step as of December 1969. The portfolio is re-balanced with initial allocations found on monthly basis. LPT is valid when the portfolios which are optimised for drawdowns provide smaller extreme drawdowns than portfolios which are optimised for variance (i.e. MPT portfolios). 

This test was carried out for all three above-mentioned security pairs for $\theta\in[1,5]$, total 15 times. In Table~\ref{table:LPT-stat}, the back-test is presented with previous example of {\em NA} and {\em P}. 
\begin{table}[loc=htbp]
\caption{The return statistics of MSCI North America / Pacific portfolios that are minimized under different definitions of risk}
\label{table:LPT-stat}
\begin{tabular}{|l|c|c|c|}
\hline
Risk type&MPT risk&Fluctuation risk&Drawdown risk\\ \hline
Min&-23.3\%&-23.4\%&-23.13\%\\ \hline
Max&14.39\%&14.1\%&14.92\%\\ \hline
Mean&0.9\%&0.9\%&0.91\%\\ \hline
Standard deviation&4.2\%&4.21\%&4.18\%\\ \hline
Mean (annualised)&10.83\%&10.79\%&10.89\%\\ \hline
Std. dev. (annualised)&14.54\%&14.59\%&14.48\%\\ \hline
\end{tabular}
\end{table}

Our results confirmed the theory: in all 15 cases the portfolio found by using LPT provided smaller absolute drawdown. Interestingly, all of the LPT portfolios had also higher return compared to those optimised for MPT.

\section{Discussion}
A simple and robust method of non-Gaussian portfolio optimisation has been 
devised. The concept of Gaussian noise kernel for the vector of daily asset return vectors led us to the separation of two types of risks. The portfolios that were designed to minimise drawdown risk provided higher average return with lower maximum drawdowns, than the 
minimum-risk portfolio according to the MPT.  
Although we have found very good results with the given data-sets, there is a clear need for further extensive analysis with real data and portfolios consisting of more than two assets. 

\begin{acknowledgments}
The support of Estonian SF grant No.\ 5036 is acknowledged. Robert Kitt also wishes to thank Hansa Investment Funds Ltd. 
\end{acknowledgments}

\bibliographystyle{apsrev}
\bibliography{kitt}

\begin{thebibliography}{12}
\expandafter\ifx\csname natexlab\endcsname\relax\def\natexlab#1{#1}\fi
\expandafter\ifx\csname bibnamefont\endcsname\relax
  \def\bibnamefont#1{#1}\fi
\expandafter\ifx\csname bibfnamefont\endcsname\relax
  \def\bibfnamefont#1{#1}\fi
\expandafter\ifx\csname citenamefont\endcsname\relax
  \def\citenamefont#1{#1}\fi
\expandafter\ifx\csname url\endcsname\relax
  \def\url#1{\texttt{#1}}\fi
\expandafter\ifx\csname urlprefix\endcsname\relax\def\urlprefix{URL }\fi
\providecommand{\bibinfo}[2]{#2}
\providecommand{\eprint}[2][]{\url{#2}}

\bibitem[{\citenamefont{Mantegna and Stanley}(2000)}]{mantegna00}
\bibinfo{author}{\bibfnamefont{R.}~\bibnamefont{Mantegna}} \bibnamefont{and}
  \bibinfo{author}{\bibfnamefont{H.}~\bibnamefont{Stanley}},
  \emph{\bibinfo{title}{An Introduction to Econophysics: Correlations and
  Complexity in Finance}} (\bibinfo{publisher}{Cambridge University Press},
  \bibinfo{address}{Cambridge}, \bibinfo{year}{2000}).

\bibitem[{\citenamefont{Voit}(2001)}]{voit01}
\bibinfo{author}{\bibfnamefont{J.}~\bibnamefont{Voit}},
  \emph{\bibinfo{title}{The Statistical Mechanics of Financial Markets}}
  (\bibinfo{publisher}{Springer-Verlag}, \bibinfo{address}{Berlin-Heidelberg},
  \bibinfo{year}{2001}).

\bibitem[{\citenamefont{Roehner}(2002)}]{roehner02}
\bibinfo{author}{\bibfnamefont{B.}~\bibnamefont{Roehner}},
  \emph{\bibinfo{title}{Patterns of Speculation: A Study in Observational
  Econophysics}} (\bibinfo{publisher}{Cambridge University Press},
  \bibinfo{address}{Cambridge}, \bibinfo{year}{2002}).

\bibitem[{\citenamefont{Bouchaud and Potters}(2003)}]{bouchaud03}
\bibinfo{author}{\bibfnamefont{J.-P.} \bibnamefont{Bouchaud}} \bibnamefont{and}
  \bibinfo{author}{\bibfnamefont{M.}~\bibnamefont{Potters}},
  \emph{\bibinfo{title}{Theory of Financial Risk and Derivative Pricing}}
  (\bibinfo{publisher}{Cambridge University Press},
  \bibinfo{address}{Cambridge}, \bibinfo{year}{2003}), \bibinfo{edition}{2nd}
  ed.

\bibitem[{\citenamefont{Muzy et~al.}(2001)\citenamefont{Muzy, Sornette, Delour,
  and Arneodo}}]{muzy01}
\bibinfo{author}{\bibfnamefont{J.}~\bibnamefont{Muzy}},
  \bibinfo{author}{\bibfnamefont{D.}~\bibnamefont{Sornette}},
  \bibinfo{author}{\bibfnamefont{J.}~\bibnamefont{Delour}}, \bibnamefont{and}
  \bibinfo{author}{\bibfnamefont{A.}~\bibnamefont{Arneodo}},
  \bibinfo{journal}{Quantitative Finance} \textbf{\bibinfo{volume}{1}},
  \bibinfo{pages}{599} (\bibinfo{year}{2001}).

\bibitem[{\citenamefont{Markowitz}(1991)}]{markowitz91}
\bibinfo{author}{\bibfnamefont{H.}~\bibnamefont{Markowitz}},
  \emph{\bibinfo{title}{Portfolio Selection}} (\bibinfo{publisher}{Blackwell
  Publishers}, \bibinfo{address}{Oxford}, \bibinfo{year}{1991}),
  \bibinfo{edition}{2nd} ed.

\bibitem[{\citenamefont{Elton and Gruber}(1995)}]{elton95}
\bibinfo{author}{\bibfnamefont{E.}~\bibnamefont{Elton}} \bibnamefont{and}
  \bibinfo{author}{\bibfnamefont{M.}~\bibnamefont{Gruber}},
  \emph{\bibinfo{title}{Modern Portfolio Theory and Investment Analysis}}
  (\bibinfo{publisher}{John Wiley \& Sons}, \bibinfo{year}{1995}),
  \bibinfo{edition}{5th} ed.

\bibitem[{\citenamefont{Gabaix et~al.}(2003)\citenamefont{Gabaix, Gopikrishnan,
  Plerou, and Stanley}}]{gabaix03}
\bibinfo{author}{\bibfnamefont{X.}~\bibnamefont{Gabaix}},
  \bibinfo{author}{\bibfnamefont{P.}~\bibnamefont{Gopikrishnan}},
  \bibinfo{author}{\bibfnamefont{V.}~\bibnamefont{Plerou}}, \bibnamefont{and}
  \bibinfo{author}{\bibfnamefont{H.}~\bibnamefont{Stanley}},
  \bibinfo{journal}{Nature} \textbf{\bibinfo{volume}{423}},
  \bibinfo{pages}{267} (\bibinfo{year}{2003}).

\bibitem[{\citenamefont{Longin and Solnik}(1998)}]{longin98}
\bibinfo{author}{\bibfnamefont{F.}~\bibnamefont{Longin}} \bibnamefont{and}
  \bibinfo{author}{\bibfnamefont{B.}~\bibnamefont{Solnik}},
  \bibinfo{type}{Tech. Rep.} \bibinfo{number}{646},
  \bibinfo{institution}{Groupe HEC} (\bibinfo{year}{1998}).

\bibitem[{\citenamefont{Longin and Solnik}(2001)}]{longin01}
\bibinfo{author}{\bibfnamefont{F.}~\bibnamefont{Longin}} \bibnamefont{and}
  \bibinfo{author}{\bibfnamefont{B.}~\bibnamefont{Solnik}},
  \bibinfo{journal}{Journal of Finance} \textbf{\bibinfo{volume}{56}},
  \bibinfo{pages}{649} (\bibinfo{year}{2001}).

\bibitem[{\citenamefont{Cizeau et~al.}(2001)\citenamefont{Cizeau, Potters, and
  Bouchaud}}]{cizeau01}
\bibinfo{author}{\bibfnamefont{P.}~\bibnamefont{Cizeau}},
  \bibinfo{author}{\bibfnamefont{M.}~\bibnamefont{Potters}}, \bibnamefont{and}
  \bibinfo{author}{\bibfnamefont{J.-P.} \bibnamefont{Bouchaud}},
  \bibinfo{journal}{Quantitative Finance} \textbf{\bibinfo{volume}{1}},
  \bibinfo{pages}{217} (\bibinfo{year}{2001}).

\bibitem[{\citenamefont{Chow et~al.}(1999)\citenamefont{Chow, Jacquier,
  Kritzman, and Lowry}}]{chow99}
\bibinfo{author}{\bibfnamefont{G.}~\bibnamefont{Chow}},
  \bibinfo{author}{\bibfnamefont{E.}~\bibnamefont{Jacquier}},
  \bibinfo{author}{\bibfnamefont{M.}~\bibnamefont{Kritzman}}, \bibnamefont{and}
  \bibinfo{author}{\bibfnamefont{K.}~\bibnamefont{Lowry}},
  \bibinfo{journal}{Financial Analysts Journal} \textbf{\bibinfo{volume}{55}},
  \bibinfo{pages}{65} (\bibinfo{year}{1999}).

\end{thebibliography}

\end{document}